\documentclass[11pt,twoside]{article}  
\usepackage{graphicx}
\usepackage{psfig}
\usepackage{epsfig}
\usepackage{adassconf}

\begin{document}   

\paperID{P.147}

\title{Search for New Open Clusters in Huge Catalogues}

\author{Ivan Zolotukhin}
\affil{Moscow State University;
       Institute of Astronomy of Russian Academy of Sciences;
}
\author{Sergey Koposov}
\affil{Sternberg Astronomical Institute; 
       Email: math@sai.msu.ru
}
\author{Elena Glushkova}
\affil{Sternberg Astronomical Institute}

%
%

\contact{Sergey Koposov}
\email{math@sai.msu.ru}

\paindex{Zolotukhin, I.}
\aindex{Koposov, S.}     
\aindex{Glushkova, E.}

\authormark{Zolotukhin, Koposov, Glushkova}


\begin{abstract}          
Current catalogues of open clusters are rather heterogeneous 
and incomplete list of clusters than true catalogues.
Before there has been no attempts of 
automatic search for open clusters in huge photometric catalogues using
homogeneous all-sky approach.

We have developed such a method based on extraction from catalogue and
successive analysis of stellar densities using the colour-magnitude diagrams.
Our algorithm finds density
peaks and then verifies the significance of these peaks exploiting the fact
that in real clusters only stars lying on the isochrone must show
density peaks, in contrast with stars lying far from the isochrone on the
CMD.
In addition such procedure allows to determine the
physical parameters: age, distance and color excess of open clusters.

Preliminary study of 150 sq. degrees in Galaxy anticenter region
yielded several dozens of new clusters. The software developed will allow
to build
first homogeneous all-sky open cluster catalogue (it will include
essentially refined data for known clusters as well) based on 2MASS data.
It will be possible to apply automated pipeline on any
multicolor catalogue, where the VO-compliant way of access (e.g. ADQL
+ SkyNode) is provided (SDSS, etc.).
\end{abstract}

\section{Introduction}

Here we present here several new methods which allow us to involve the data
from existing huge stellar catalogues into search and analysis of
star clusters. We also show some preliminary but very promising
results of these methods.

Currently there are several compiled catalogues of star clusters,
but they all have several significant disadvantages:

\begin{itemize}
\item
Tens of percents of their clusters do not have reasonable CMD measurements, so 
they are only unverified groups of stars
\item
Some of them do not even have precise celestial coordinates
measurements.
\item
Even some rich clusters do not have important physical parameters measured
(age, distance, color excess)
\item
In general, all existing catalogues of open clusters are very 
heterogeneous and are just the compilations of data from different sources.
Nobody has ever performed fully automatic homogeneous search of clusters 
in full-sky regime.

\end{itemize}

\section{Automated Algorithm}

Now uniform data from huge stellar catalogues allow to apply
homogeneous approach to the search of new and investigation of all
known open clusters. Our method consists of two main procedures:

\begin{itemize}
\item
  search for open clusters based on stellar density analysis
\item
  verification if given peak of density is a real cluster but not an
  occasional group of field stars 
\item estimation of main parameters of clusters
\end{itemize}

First procedure is not as trivial as simple analysis of surface
density of objects from catalogue because of strongly variable mean
("background") stellar density from place to place. So one can not
apply a constant threshold filter to get a list of density peaks that
would be the open cluster candidates. It is necessary to use a special
convolution function to distinct peaks from such a complex
background. But still we can not be sure that density peaks
detected are not occasional groups of stars. We need to prove
 that stars in these groups are physically related with,
so that they should lie on the same isochrone.

\begin{figure}
\epsscale{.70}
\plotone{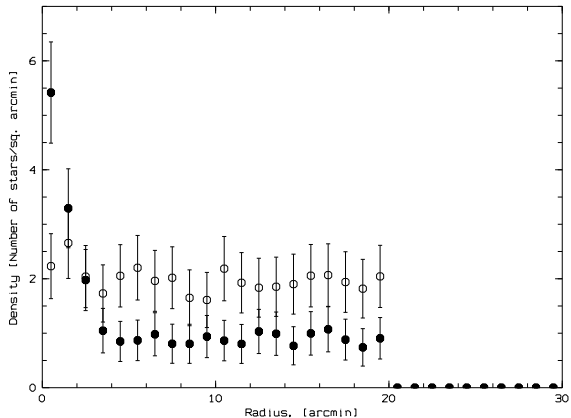}
\caption{The radial density distribution of stars lying on fitted isochrone
(filled circles) and that of field stars, lying far from
the isochrone (empty circles) for the cluster KSE 18. The peak of the
cluster density is clearly seen, whereas the density of field stars is strictly
constant.}
\label{kse18_density}
\end{figure}

\begin{figure}
\epsscale{.40}
\plotone{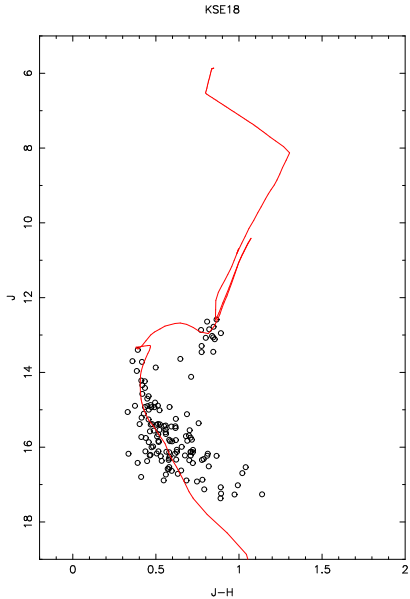}
\caption{The color-magnitude diagram of the new cluster KSE 18 
($RA = 05^h 53^m 48.9^s, Dec = +26^{\deg} 50' 25''$) with the fitted isochrone
of age 600 Myr and distance of 3.1 kpc. }
\label{kse18_isochrone}
\end{figure}

Our algorithm is based on the fact that real cluster members lie near
the isochrone and show the density peak, whereas field stars must demonstrate
flat distribution (Fig.~\ref{kse18_density}). It allows us to find the
position of the isochrone of the cluster even when CMD is
"polluted" by field stars. Such a fitting by isochrone yields
cluster's distance, age and color excess (Fig.~\ref{kse18_isochrone}).

\section{Preliminary Results}

These two algorithms are developed using C language and tested on
several fields from 2MASS catalogue which is the primary catalogue of
our interest. From the very preliminary analysis of the 150
sq. degrees in the Galaxy anticenter region we have found several
dozens of new clusters, and determined parameters for 10 of them. 
In addition we have determined parameters for several
previously known open clusters for which parameters have not been 
determined before.

\begin{figure}
\epsscale{.80}
\plottwo{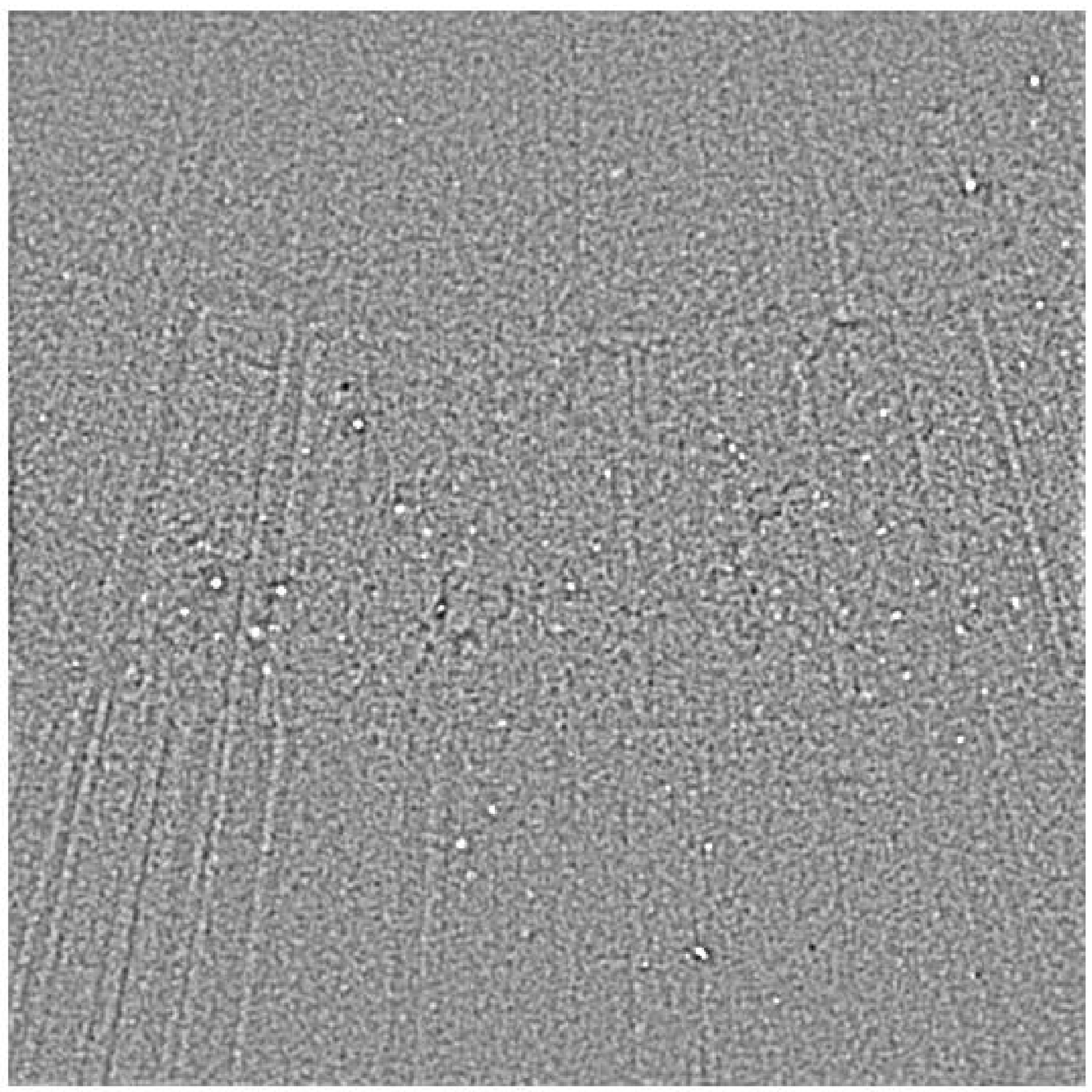}{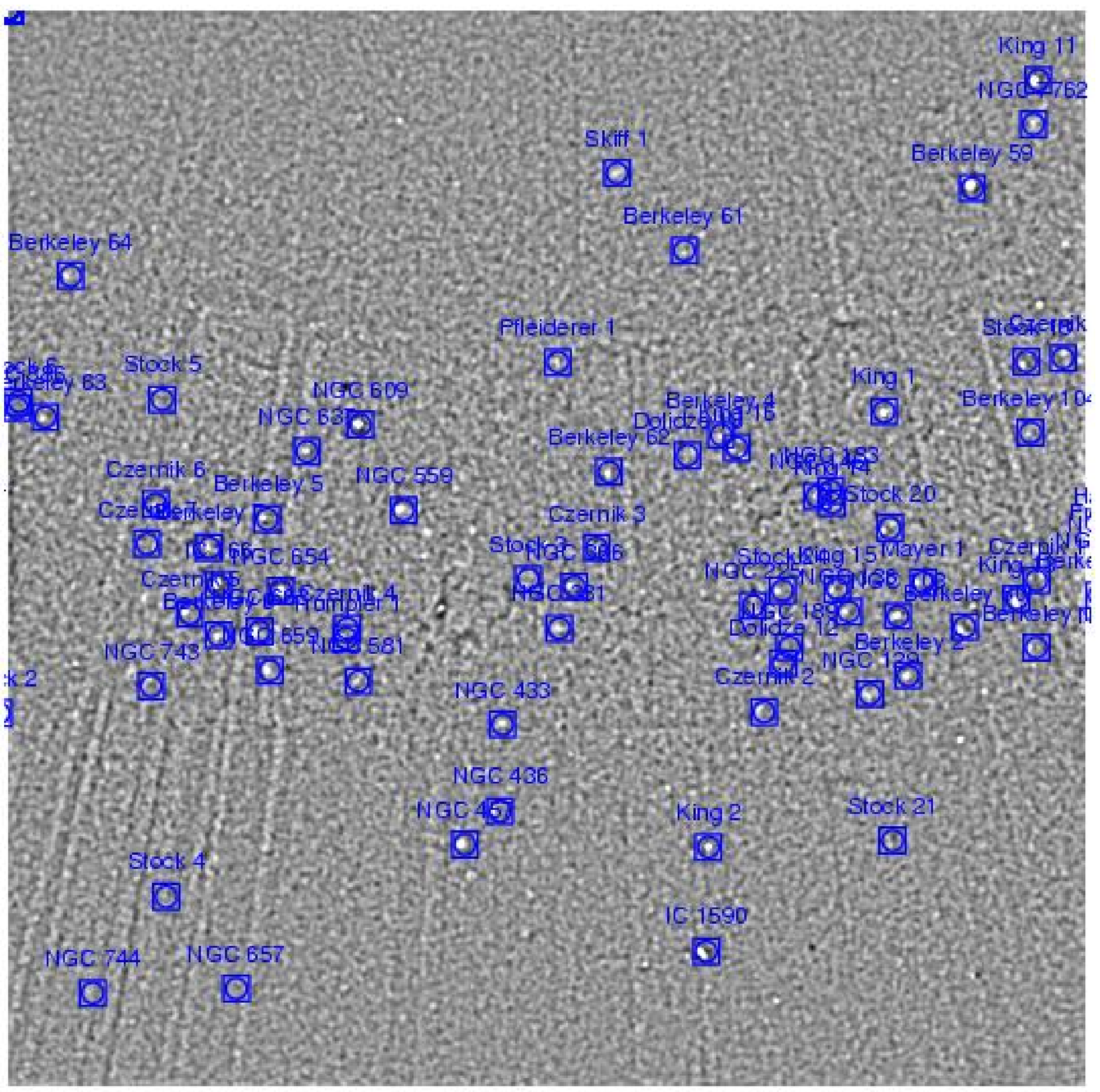}
\caption{Perseus arm region ($16^\circ \times 16^\circ$). The left
panel is the overdensity map obtained by our methods, the right panel
is the same map with superimposed known clusters. It can be seen
that most of the known clusters correspond to the peaks on our map, so
they can be easily detected by our method. Also, there are several
peaks that do not coincide with known clusters. These are new
clusters.}
\label{perseus}
\end{figure}

Also, a set of new clusters was found when we looked in the
region of Perseus arm (Fig.~\ref{perseus}). It should be noticed, that
the clusters which we have found are not infrared clusters, most
of them are clearly visible in the optical wavelength range.

So finally, we can find clusters, confirm their reality, and 
reliably determine their main parameters (age, distance and color excess). Our
main goal is to process all-sky selection from 2MASS stellar catalogue
and make up first homogeneous catalogue of open clusters which will
include newly discovered open clusters and already known ones
with their parameters determined by our uniform approach. Also it
would be promising to include catalogues from other surveys, such as
DENIS and SDSS, into open clusters search.

\acknowledgments IZ wish to thank Institute of Astronomy of Russian
Academy of Sciences and OFN RAS program ``Extended Objects in the
Universe'' for the ADASS conference attendance support.


\end{document}